\documentclass{article}
\PassOptionsToPackage{numbers, compress}{natbib}
\usepackage[preprint]{neurips_2025}
\usepackage[utf8]{inputenc} 
\usepackage[T1]{fontenc}    
\usepackage{hyperref}       
\usepackage{url}            
\usepackage{booktabs}       
\usepackage{amsfonts}       
\usepackage{nicefrac}       
\usepackage{microtype}      
\usepackage{amsmath}
\usepackage{xcolor}         
\usepackage[pdftex]{graphicx}
\usepackage{wrapfig}
\usepackage{chngcntr} 
\usepackage{xspace}

\title{Measuring Dependencies between Biological Signals with Self-supervision, and its Limitations}
\workshoptitle{AI for Science Workshop}

\makeatletter
\newcommand{\ie}{i.e.\@ifstar{\xspace}{,\xspace}} 
\newcommand{\Ie}{I.e.\@ifstar{\xspace}{,\xspace}} 
\newcommand{\eg}{e.g.\@ifstar{\xspace}{,\xspace}} 
\newcommand{\Eg}{E.g.\@ifstar{\xspace}{,\xspace}} 
\makeatother
\author{%
\begin{tabular}{c}
Evangelos Sariyanidi\textsuperscript{1}\thanks{sariyanide@chop.edu}, John D. Herrington\textsuperscript{1,2}, Lisa Yankowitz\textsuperscript{1}, %
 Pratik Chaudhari\textsuperscript{2}, \\Theodore D. Satterthwaite\textsuperscript{2},
Casey J. Zampella\textsuperscript{1}, Robert T. Schultz\textsuperscript{1,2}, \\%
 Russell T. Shinohara\textsuperscript{2},
 Birkan Tunc\textsuperscript{1,2} \\[2pt]
\textsuperscript{1}The Children's Hospital of Philadelphia \quad \textsuperscript{2}University of Pennsylvania 
\end{tabular}
}

\begin{document}

\maketitle

\begin{abstract}
Measuring the statistical dependence between observed signals is a primary tool for scientific discovery. However, biological systems often exhibit complex non-linear interactions that currently cannot be captured without a priori knowledge regarding the nature of dependence. We introduce a self-supervised approach, concurrence, which is inspired by the observation that if two signals are dependent, then one should be able to distinguish between temporally aligned vs.\ misaligned segments extracted from them. Experiments with fMRI, physiological and behavioral signals show that, to our knowledge, concurrence is the first approach that can expose relationships across such a wide spectrum of signals and extract scientifically relevant differences without ad-hoc parameter tuning or reliance on a priori information, providing a potent tool for scientific discoveries across fields. However, depencencies caused by extraneous factors remain an open problem, thus researchers should validate that exposed relationships truely pertain to the question(s) of interest.
\end{abstract}

\section{Introduction}
Measuring dependencies between biological signals is fundamental for understanding the complex interplay within and between molecular, neurobiological, and behavioral processes. The most common approach to quantifying statistical dependence is using linear model-based statistics, with the Pearson correlation coefficient being the dominant metric \cite{tjostheim22}. However, biological systems often exhibit interactions \cite{janson12} that cannot be captured by linear models \cite{he21}, such as cross-frequency coupling \cite{schmidt97,hyafil15,yeh23}, threshold effects \cite{beltrami95}, phase shifts~\cite{tiesinga10}, feedback systems \cite{beltrami95,cosentino11}, or multi-scale interactions~\cite{qu11}. 

While linear models cannot comprehensively capture statistical dependence, linear and non-linear models together can. That is, if two time signals x and y are dependent but uncorrelated, then there must be (non-linear) mathematical transformations $f$ and $g$ such that the transformed signals $f(x)$ and $g(y)$ are correlated~\cite{renyi05}. However, the specific transformations that expose the dependence can be particular to each problem and difficult to identify when the compared signals are generated by complex or unknown mechanisms. The Hilbert-Schmidt Independence Criterion~\cite{gretton07} –which can be considered as a generalization of distance correlation~\cite{szekely07,sejdinovic13}– or variants of canonical correlation analysis~\cite{verbeek03,andrew13} can, in principle, determine linear and non-linear dependence. However, these approaches are successful only if one can identify model parameters or kernels that expose the dependence~\cite{hua15,gretton12}, which may not be possible or may require large samples~\cite{zhuang20,marek22}. Alternatively, one may use analytical transformations such as Fourier or wavelet decomposition \cite{greenblatt12,fujiwara16,schmidt12}, but the generalizability of this approach is limited, as there is no single transformation that works for all signals \cite{mallat09,vetterli14}. Moreover, analytical transformations pose family-wise error problems \cite{maraun04,kramer08} because they typically decompose each signal into multiple signals (e.g., frequency bands), and dependence can occur between any pair of decomposed signals (e.g., cross-frequency dependence). These issues are exacerbated when the compared signals are multi-dimensional and only a subset in one set of signals depends on an unknown subset in the other. In sum, currently there is no tractable method that can detect or quantify the dependence between a broad variety of biological signals when the dependence structure is not known a priori—presenting a major obstacle to scientific discovery.

\begin{figure}
\includegraphics[width=\textwidth]{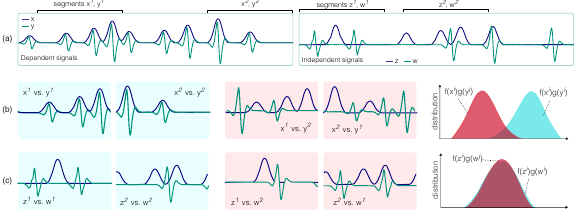}
\caption{(a) Dependent signals $x$ and $y$, where $x$ resembles the integral of $y$; and a pair of independent signals $w$ and $z$. (b) Concurrent segments from dependent signals $x$ and $y$  have different characteristics from non-concurrent segments, therefore it is possible to find functions $f$ and $g$ such that when they are respectively applied to concurrent segments of $x$ and $y$ they produce a larger inner product compared to when they are applied to non-concurrent segments (e.g., $f$ as the identity operator and $g$ the integral operator. (c) The concurrent and non-concurrent segments from independent signals exhibit similar characteristics under the $f$ and $g$ that separate dependent segments.}
\label{fig:main}
\end{figure}

We introduce a new approach, called concurrence, to quantify the statistical dependence between pairs of signals. The proposed approach is built on the following heuristic: if two signals are statistically dependent, then  temporally aligned (i.e., concurrent) segments of the compared signals must be separable from segments that are temporally misaligned (i.e., not concurrent), as illustrated in Fig.~\ref{fig:main}. 
This separability criterion provides a straightforward recipe for automatically finding linear or non-linear transformations that expose dependence—namely, training a machine learning model that classifies between concurrent vs. non-concurrent segments extracted from signals (Fig.~\ref{fig:main}). We test this approach in experiments with three distinct biological signal types (i.e., fMRI, physiological, and behavioral data). Results indicate that the concurrence approach can expose a large range of dependencies without any ad-hoc modification, as the same parameters successfully detected dependence in all three signal types; and can handle noise and stochastic dependence (Fig.~\ref{fig:snr}). Furthermore, our approach yields a score coined the concurrence coefficient, which is scaled between 0 and 1. 
Finally, concurrence is computationally efficient and we provide an open-source software package to implement it for research purposes. 

\section{The Concurrence Approach}

Biological signals often contain dependencies that can be observed within relatively short time windows. For example, the dependence between respiration and cardiac activity occurs regularly and can be observed within a few seconds \cite{russo17}. Behavioral signaling during conversations contains events indicative of social coordination, such as mimicry \cite{lakin03} and backchanneling \cite{shelley13}, which also occur within a short time window. Dependencies between activity in different parts of the brain can be observed (e.g., using fMRI) instantaneously or within a short time delay, usually milliseconds. As such, one can focus on relatively small time chunks and still capture a broad range of dependencies.

Suppose that $x_{t,w}$ and $y_{t,w}$ are segments of signals $x$ and $y$, observed between the time points $t$ and $t+w$. If both $x_{t,w}$ and $y_{t,w}$ contain (finite) responses to a common event, 
then they must be statistically dependent. Thus, there must exist transformations $f$ and $g$ such that transformed versions of the segments, $f(x_{t,w})$ and $g(y_{t,w})$, are correlated \cite{renyi05}. The crux of our approach is that, while $f(x_{t,w})$ and $g(y_{t,w})$ are expected to be correlated,  $f(x_{t,w})$  and $g(y_{t',w})$ are, on average, uncorrelated if $t'$ is a random time point, different from $t$. For example, if $x$ and $y$ are behavioral signals of two people in a conversation, $t$ may indicate a time when one partner spontaneously mimics the behavior of the other, and it is unlikely that the same mimicry pattern is present at a random time $t'$. 


Since extracting concurrent or non-concurrent segments from synchronized pairs of signals is trivial, this intuitive idea provides a straightforward manner of detecting linear or non-linear dependence fully automatically, without relying on an appropriate choice of kernel or other hyperparameters. Specifically, we quantify dependence through the concurrence coefficient, which is obtained by training a machine learning model to classify between randomly cropped concurrent versus non-concurrent segments in a dataset of signal pairs, and then calculating the normalized classification accuracy on another dataset $D= {(x^1,y^1), (x^2,y^2), … (x^N,y^N)}$ as:
\begin{equation}
\text{concurrence coefficient} = 2 \times \max (\text{accuracy}, 0.5)-1
\label{eq:cc}
\end{equation}
This dataset $D$ must not overlap with the dataset used while training the machine learning model, lest the classifier may overfit and overestimate the dependence. Thus, one may use a cross-validation (CV) procedure and compute the average concurrence coefficient over the test sets of the CV folds, or compute the concurrence coefficient on an independent dataset.

The concurrence coefficient is bounded between 0 and 1, and its magnitude indicates the strength of dependence between the compared signal pairs (Fig.~\ref{fig:snr}). 

\subsection{The per-segment concurrence score}
The classifier that we use for distinguishing between concurrent and non-concurrent segments is a neural network that is trained to produce a \textit{per-segment concurrence score} (PSCS) $s$ from a given pair of segments $x_{t,w}$ and $y_{t',w}$, such that $s > 0$ if $t=t'$, and $s \leq 0$ if $t\neq t'$. The PSCS is not only used to compute the concurrence coefficient on a dataset of signals $D$ as in~\eqref{eq:cc}, but it also allows one to quantify the dependence between a specific pair of segments. As such, the concurrence coefficient and the PSCS have two distinct uses for scientific analyses. While the concurrence coefficient can uncover whether and to what extent two biological processes (e.g., breathing rate and cardiac activity) are related in general (i.e., at the sample level), the PSCS between concurrent segments (i.e., $t=t'$) can uncover whether this relationship is stronger for a specific individual, for individuals with a certain condition (e.g., anxiety), or for certain moments within the compared signals. Our experiments on real data include use cases for both the concurrence coefficient and the PSCS.

We compute the PSCS through a three-layer classifier that enables the interpretation of the results. Suppose the segments $x_{t,w}$ and $y_{t',w}$ are extracted respectively from $K_x$- and $K_y$-dimensional signals. Then, we first transform these segments through separate functions $f$ and $g$ into segments of dimensions $K_f$ and $K_g$, 
\begin{align}
f : \mathbb{R}^{K_x \times w} \to \mathbb{R}^{K_f \times w'},\,\, g : \mathbb{R}^{K_y \times w} \to \mathbb{R}^{K_g \times w'}
\end{align}
where $w'$ is the temporal length of the transformed segments. Then, we compute the covariance $C = f(x_{t,w}) g(y_{t',w})^T$, where $T$ is the transpose operator. Then we finally compute the PSCS through a linear layer,
\begin{equation}
 s = \sum_{i} \sum_{j}  \alpha_{ij} C_{ij},
\end{equation}
where $C_{ij}$ is the $ij$th entry of $C$, and the corresponding $\alpha_{ij}$ are the learned weights. As such, the PSCS $s$ can simply be considered to be the weighted average of the covariance entries between the transformed segments, where the transformations $f$, $g$ and the weights are learned while training this network to separate between concurrent and non-concurrent segments.

\subsection{Implementation}

To make the concurrence approach useful for scientific purposes, the functions $f$ and $g$ should be flexible enough to expose arbitrary dependencies. Also, one needs a training procedure that requires no hyperparameter tuning and can work successfully even with modestly-sized datasets, since the samples used in scientific analyses often have only hundreds or even fewer samples. 

As such, we model the transformations $f$ and $g$ with Convolutional Neural Networks (CNNs), which are universal approximators \cite{zhou20} and, thanks to advances in machine learning \cite{wang16,bai18}, have well-established recipes for training across a large variety of temporal analysis tasks without ad-hoc modifications~\cite{wang16}, particularly when modeling short-term dependencies \cite{bai18}. Our experiments with real and synthetic data verify that CNNs with the \textit{same parameters} (Table~\ref{tab:params}) can detect a wide range of linear or non-linear dependence patterns between signals that have distinct frequency characteristics and are corrupted by large amounts of noise. Further, the training does not require an unrealistic sample size, as our experiments show that fewer than 100 signal pairs can suffice.

\subsection{The effect of the segment size}
\label{sec:w}
The only parameter that needs to be determined by the user is the segment size $w$. Fortunately, setting this parameter is not difficult, as one can err on the side of a large $w$ value without risking to miss a potential dependence, as explained in this section. 

If one picks a $w$ value that is too small, the resulting segments can fail to include the activated portions of both signals, or only one of them (e.g., when there is a time lag). On the other hand, picking larger $w$ values should, in principle, not lead to a missed dependence, since the longer segment, the more the information to observe a potential dependence. 

\begin{figure}
\centering
\includegraphics[width=0.84\textwidth]{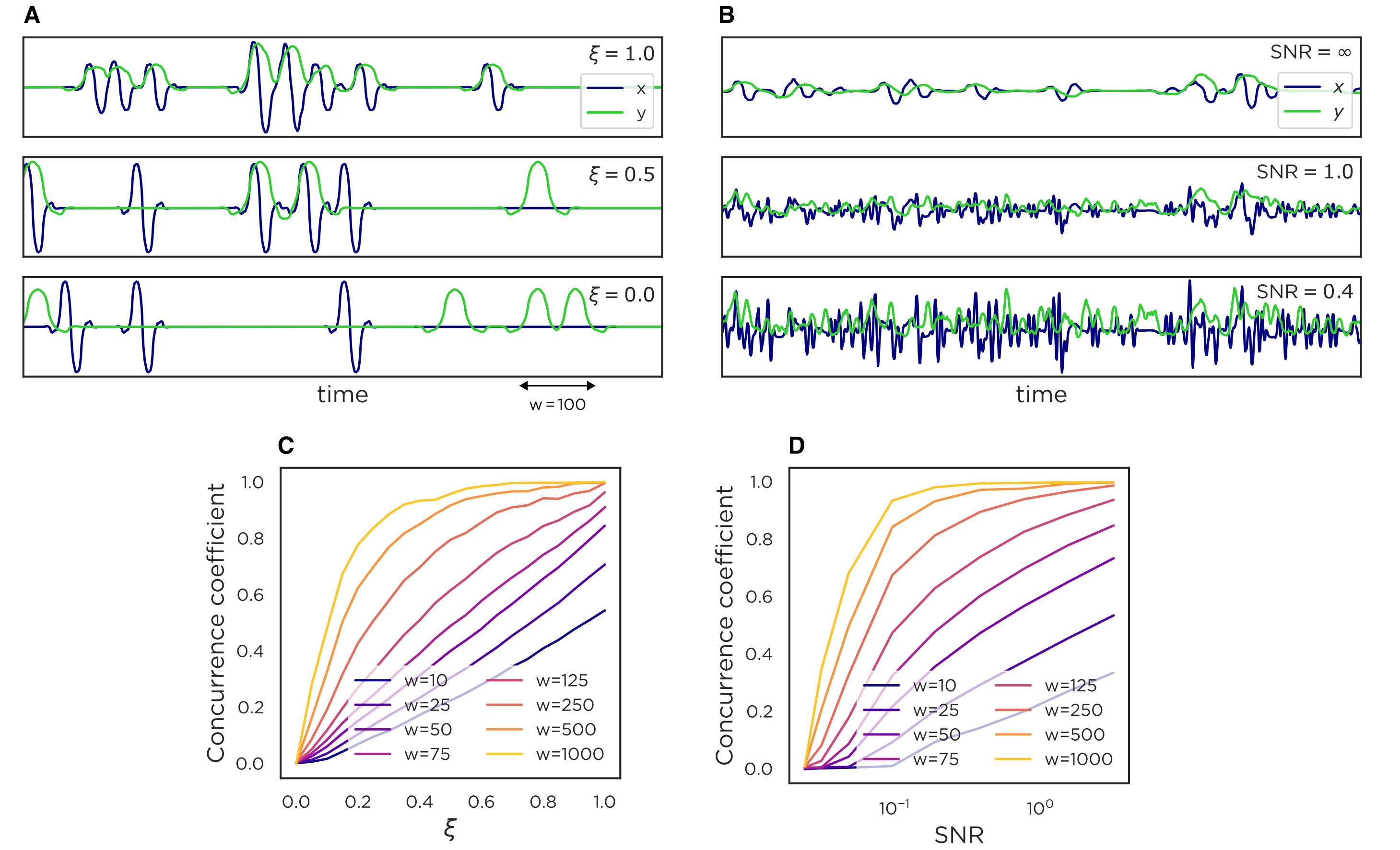} 
\caption{Synthesized signals with various structures of dependence that can be detected with the concurrence coefficient. (A) Examples of synthesized signals with deterministic dependence ($\xi=1.0$), stochastic dependence ($\xi=0.5$) and no dependence ($\xi=0.0$). (B) Dependent pairs of signals with varying degrees of noise.  (C) Concurrence coefficient vs.\ degree of dependence (i.e., $\xi$). (D) Concurrence coefficient vs. signal-to-noise ratio (SNR).}
\label{fig:snr}
\end{figure}

To empirically validate this intuitive expectation, we compute concurrence from signals $x$ and $y$ generated as realizations of random sequences (RSs) $\mathbf x[t]$ and $\mathbf y[t]$ with controlled dependence. Specifically, $\mathbf x[t]$ and $\mathbf y[t]$ are obtained by convolving impulse trains $\mathbf h_x[t]$ and $\mathbf h_y[t]$ through (deterministic) filters $\phi$ and $\psi$, with added noise $\mathbf n_x$ and $\mathbf n_y$:
\begin{align}
\mathbf x[t] &= (\phi \star \mathbf{h}_x)[t]+ \sigma_{\text{n}x} \mathbf n_x[t] \label{eq:modelx}
\\
\mathbf y[t] &= (\phi \star \mathbf{h}_y)[t]+ \sigma_{\text{n}y} \mathbf n_y[t].
\label{eq:modely}
\end{align}
The noise processes $\mathbf n_x$ and $\mathbf n_x$  are independent of each other, and the $\sigma_{\text{n}x}$ and $\sigma_{\text{n}y}$ parameters control the  noise amount. The impulse processes $\mathbf h_x$ and $\mathbf h_y$ are modeled as:
\begin{align}
\mathbf{h}_x &= \xi \mathbf c[t] + (1-\xi)\mathbf p_x[t] \\
\mathbf{h}_y &= \xi \mathbf c[t] + (1-\xi)\mathbf p_y[t],
\end{align}
where $\mathbf c$ is the part of the impulse signal that is common between $\mathbf h_x$ and $\mathbf h_y$. The RSs $\mathbf p_x$ and $\mathbf p_y$ are the impulse processes that are independent from each other. The processes $\mathbf c$, $\mathbf p_x$ and $\mathbf p_y$ are all modeled as  Bernoulli processes. The parameter $\xi$ determines the degree of dependence between $\mathbf x$ and $\mathbf y$; the larger the $\xi$ the stronger the dependence (Fig.~\ref{fig:snr}). 

Fig.~\ref{fig:snr}C shows the concurrence coefficients obtained from pairs of signals with varying degrees of dependence (i.e., $\xi$) but without noise ($\sigma_{\text{n}x}=\sigma_{\text{n}y}=0$). Results show that the concurrence coefficient exposes dependencies across a large range of $\xi$ values. Of note, the concurrence coefficient is approximately linearly proportional to $\xi$ when $w$ is large enough to contain the entire event (e.g., $w$=100 for signals as in Fig.~\ref{fig:snr}A) but not much larger. While the degree of dependence $\xi$ is overestimated with larger segments, this does not imply that the concurrence coefficient risks detecting spurious relationships (false positives), as the concurrence coefficient is approximately zero when $\xi=0$, regardless of the segment size $w$. In sum, the exact $w$ value is of little concern when the goal is to uncover whether two processes are dependent or not, and choosing a larger $w$ value does not lead to missed or spurious dependencies.

Fig.~\ref{fig:snr}D shows that a large $w$ value is also not problematic in the presence of noise ($\sigma_{\text{n}x}>0$, $\sigma_{\text{n}y}>0$). The concurrence coefficient decreases with the signal-to-noise (SNR), yet it can uncover dependence even when the SNR is 0.10, which is lower than a worst-case estimate for fMRI data (SNR=0.35)~\cite{welvaert13}. The pattern in Fig.~\ref{fig:snr}D  is similar to that in Fig.~\ref{fig:snr}C\----increasing the segment size $w$ does not lead to missed dependencies, confirming that erring on the side of a large $w$ value is a useful strategy for exposing dependence in cases where there is no information to determine the right $w$ value.

\section{Experimental Validation}

We first validated concurrence on 100 synthetic but challenging datasets with controlled dependence, and compared it with eight methods (Section~\ref{sec:synthetic}) Next, we applied concurrence on three types of real data with single- and multi-dimensional signals and included cases of linear or non-linear dependence. Specifically, we experimented on brain imaging (fMRI), physiological (breathing and heart rate), and behavioral (facial expressions and head movements) signals.

\subsection{Experiments on Synthetic Data}
\label{sec:synthetic}
We generated 100 synthetic datasets, where each dataset contained pairs of statistically dependent signals. The goal was to determine the dependence in as many datasets as possible by using only the off-the-shelf implementation of our algorithm, without any ad-hoc parameter adjustment. The datasets were designed to be challenging, with dependencies difficult to visually ascertain~(Fig.~\ref{fig:signals},~\ref{fig:corrs}). 

\textbf{Datasets.} Each of the 100 synthesized datasets is comprised of 500 pairs of signals $(x,y)$ generated as realizations of the RSs in \eqref{eq:modelx} and \eqref{eq:modely}. The pairs were statistically dependent through a random $\xi$ value such that $0.1 < \xi \leq 1$. To increase the challenge, we made two modifications compared to the procedure described in Section~\ref{sec:w}. First, the impulse processes $\mathbf c$, $\mathbf p_x$ and $\mathbf p_y$ were made non-stationary. Specifically, the probability of observing an impulse at any of these RSs was increasing or decreasing linearly at a random rate. The second challenge was adding a random a lag by (circularly) shifting the generated $y$ signal through a random lag between 0 and 50 time frames. The convolution kernels $\phi$ and $\psi$ were determined by randomly picking a kernel and a scale from the `pywavelets` library. The noise processes $\mathbf n_x$ and $\mathbf n_y$ were also generated by convolving randomly selected kernels with separate and independent impulse processes. 

\textbf{Compared methods}. We compared concurrence with correlation (Pearson's $r$), windowed cross-correlation (WCC)~\cite{boker02}, distance correlation (DC)~\cite{szekely07}, Hilbert-Schmidt Independence Criterion (HSIC)~\cite{gretton07},  Mutual Information (MI), Conditional MI (CMI), Multiscale Graph Correlation (MGC), Kernel Mean Embedding Random Forest (KMERF). HSIC, MGC and KMERF have been implemented via the \texttt{hyppo} software package; DC was implemented via \texttt{dcor}~\cite{dcor1}; and Pearson's $r$ was implemented through \texttt{scikit-learn}. We provided our own implementation for the remaining methods. The statistical significance for all methods have been computed via permutation tests.

\textbf{Results}. Table~\ref{tab:results} provides the results of experiments on synthesized datasets. CMI is the best among methods alternative to concurrence, due probably to its ability to model non-linear dependence and temporal dependence. Still, this method can detect the dependence in only 34\% of the datasets. CMI or other alternative methods can possibly detect the dependence in more datasets if their parameters are optimized for each dataset. However, this is often not possible in the context of scientific analyses with modestly sized datasets, as one should do multiple tests correction~\cite{armstrong14} for the tested parameter values, leading to significant decrease in statistical power. Concurrence detected the dependence in 97\% of datasets, using identical network hyperparameters (Table~\ref{tab:params}) and segment size $w$.

\begin{table}
\centering
\caption{Results from the 100 synthetic datasets: The number of datasets that have been (correctly) identified as statistically dependent (at significance level 0.05) by each of the compared methods.}
\begin{tabular}{|c|c|c|c|c|c|c|c|c|c|c|} \hline
Pearson's $r$ & WCC & DC & HSIC & MI & CMI & MGC & KMERF & Concurrence \\ \hline
8 & 10 & 12 & 10 & 7 & 34& 9 & 	11 & 97\\ \hline
\end{tabular}
\label{tab:results}
\end{table}

\subsection{Applications to Real Biological Signals}
\label{sec:real}

\begin{figure}
\includegraphics[width=0.9\textwidth]{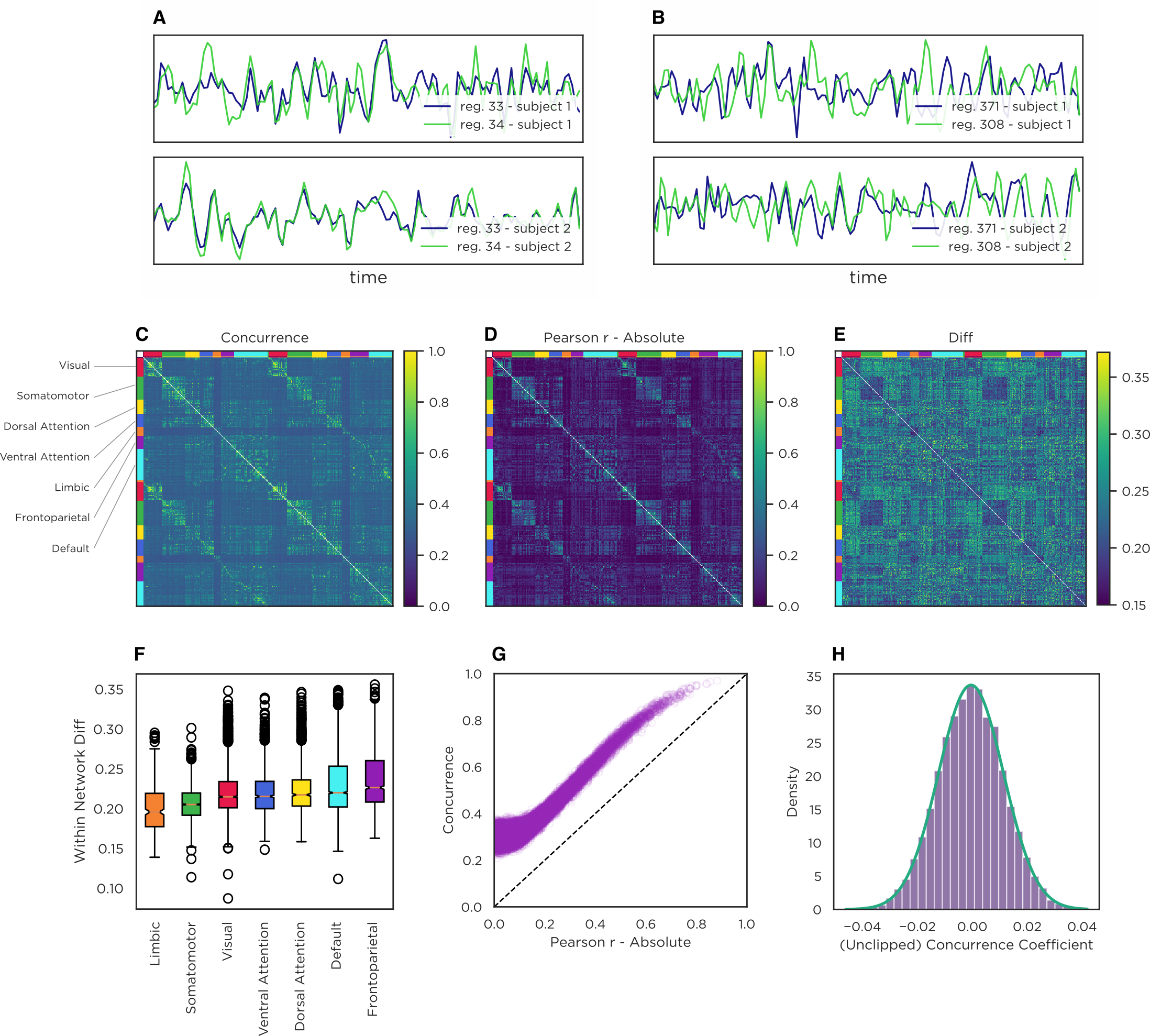} 
\caption{ (A) Signals from two brain regions with correlated fMRI signals. (B) Signals from two brain regions with fMRI signals that are dependent (concurrence coefficient: 0.25) but approximately uncorrelated (Pearson’s r: 0.02). (C) The connectivity matrix computed with the concurrence coefficient. (D) The connectivity matrix computed with (absolute) correlation values (Pearson’s r). (E) The difference between the two concurrence- and correlation-based connectivity matrices. (F)  The distribution of the difference between the concurrence- and correlation-based connectivity matrices, shown separately for the seven brain networks. (G) Alternative comparison of the Pearson’s r vs.\ concurrence coefficients computed from all the brain region pairs. (H) The (unclipped) concurrence coefficient between 10,000 pairs of brain regions of mismatched participants.}
\label{fig:fmri}
\end{figure}

\textbf{Brain Imaging}
Our experiments on fMRI signals aim to identify how strongly different brain regions are functionally connected. Pearson's $r$ is the single-most commonly used metric for this purpose \cite{liu24}. Fig.~\ref{fig:fmri} compares the connectivity matrices obtained with Pearson's $r$ (i.e., correlation matrix) and the concurrence coefficient (i.e., concurrence matrix) on a version of the Philadelphia Neurodevelopmental Cohort dataset \cite{baum20} that was pre-processed as in prior work \cite{satterthwaite16}. This dataset uses the parcellation scheme that divides each brain into 400 regions \cite{schaefer18}. The concurrence coefficient is computed on segments of size $w=30$ time points, which corresponds to approximately 90 seconds, whereas the entire signals included 120 time points. Thirty percent of the dataset (426 participants) was used to train the neural networks needed for the concurrence coefficients, and the results in both connectivity matrices were computed from the remaining 70\%.

The overall similarity between the two connectivity matrices (Fig.~\ref{fig:fmri}C vs. Fig.~\ref{fig:fmri}D) is striking and suggests that the concurrence coefficient uncovers a dependence structure that has been validated in the field. Fig.~\ref{fig:fmri}G shows that there are no pairs of regions with a concurrence score less than 0.2, even though there are many pairs that are uncorrelated (i.e., Pearson r $\approx$ 0), suggesting that concurrence captured statistical dependencies that cannot be captured with correlation (e.g., Fig.~\ref{fig:fmri}B) as well as those that can (e.g., Fig.~\ref{fig:fmri}A). The fact that the concurrence coefficient exposed a dependence between all 400×199=79,600 pairs of brain regions with 79,600 independently trained networks verifies that the training needed for the concurrence coefficient can be done robustly. We ran a permutation test to identify if the method detects spurious dependence (Type I error) by computing concurrence between signals of mismatched participants. The concurrence coefficient was closely distributed around zero (Fig.~\ref{fig:fmri}H), indicating no spurious relationships.
The differences between the concurrence coefficient and Pearson’s correlation exhibit a structured pattern across the seven brain networks (Fig.~\ref{fig:fmri}F), increasing progressively from lower-order affective (limbic), somatomotor and sensory (visual) networks to higher-order cognitive control (ventral attention, dorsal attention, default mode, frontoparietal) networks. This systematic increase suggests that linear correlation may not be capturing complex connectivity patterns that involve integrative processing or dynamic modulation. 

\begin{figure}
\includegraphics[width=\textwidth]{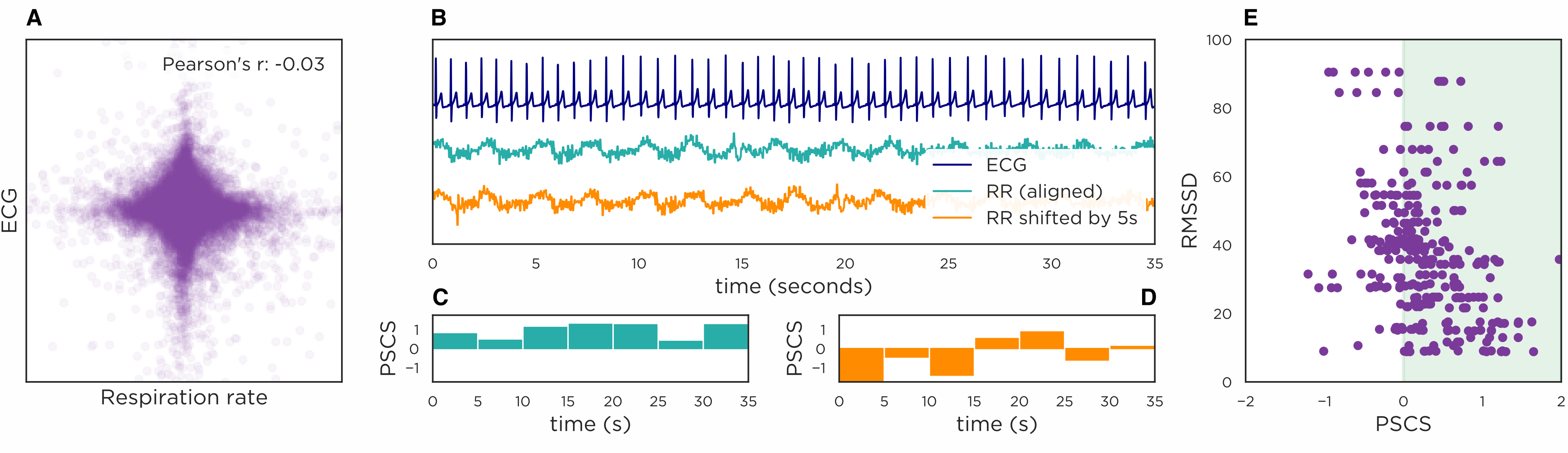} 
\caption{(A) Scatter plot and correlation between the respiration rate (RR) and ECG signal. (B) A sample ECG signal plotted against the synchronized (i.e., time-aligned) RR signal and the temporally misaligned RR. (C) The per-segment concurrence scores (PSCSs) between the temporally aligned ECG and RR are positive, which indicate that the PSCS correctly predicts that the segments are temporally aligned. (D) The PSCSs between the temporally misaligned ECG and RR are generally negative. (E) The PSCS for (temporally aligned) ECG and RR signals against the RMSSD, computed on a dataset of 30 participants for multiple segments per participant.}
\label{fig:hr}
\end{figure}

\textbf{Physiological data}.
We next investigate dependencies in a dataset of breathing and cardiac activity. While these two processes are known to be biologically linked \cite{adrian32}, the correlation between respiration rate and electrocardiogram (ECG) signals is approximately zero (Fig.~\ref{fig:hr}A). We applied the proposed method to a dataset of 60 pairs of temporally synchronized ECG and respiration rate signals, collected at the Children’s Hospital of Philadelphia using Zephyr BioModule sensors. The duration of the tasks used for data collection ranged from 4 to 7 minutes. The data were split into four subject-independent cross-validation folds. The segment size w was equivalent to 5 seconds.

The average concurrence coefficient on the test folds was 0.50 (p<0.001), indicating that the concurrence approach successfully detects the relationship between respiration rate and ECG signals. The PSCS can generally distinguish between compared signals that are temporally aligned or not (Fig.~\ref{fig:hr}B–D), validating that concurrence can identify relationships (or lack thereof) that are difficult to determine visually. Fig.~\ref{fig:hr}E plots the PSCSs from temporally aligned segments vs. the root mean square of successive differences (RMSSD) derived from the ECG signal of each interaction. That the PSCS is generally larger when the RMSSD is low may suggest that the trained algorithm predicts a stronger relationship between ECG and respiration rate when the latter is increased.

\begin{wrapfigure}{r}{0.65\linewidth} 
  \centering
  \vspace{-\baselineskip} 
\includegraphics[width=0.65\textwidth]{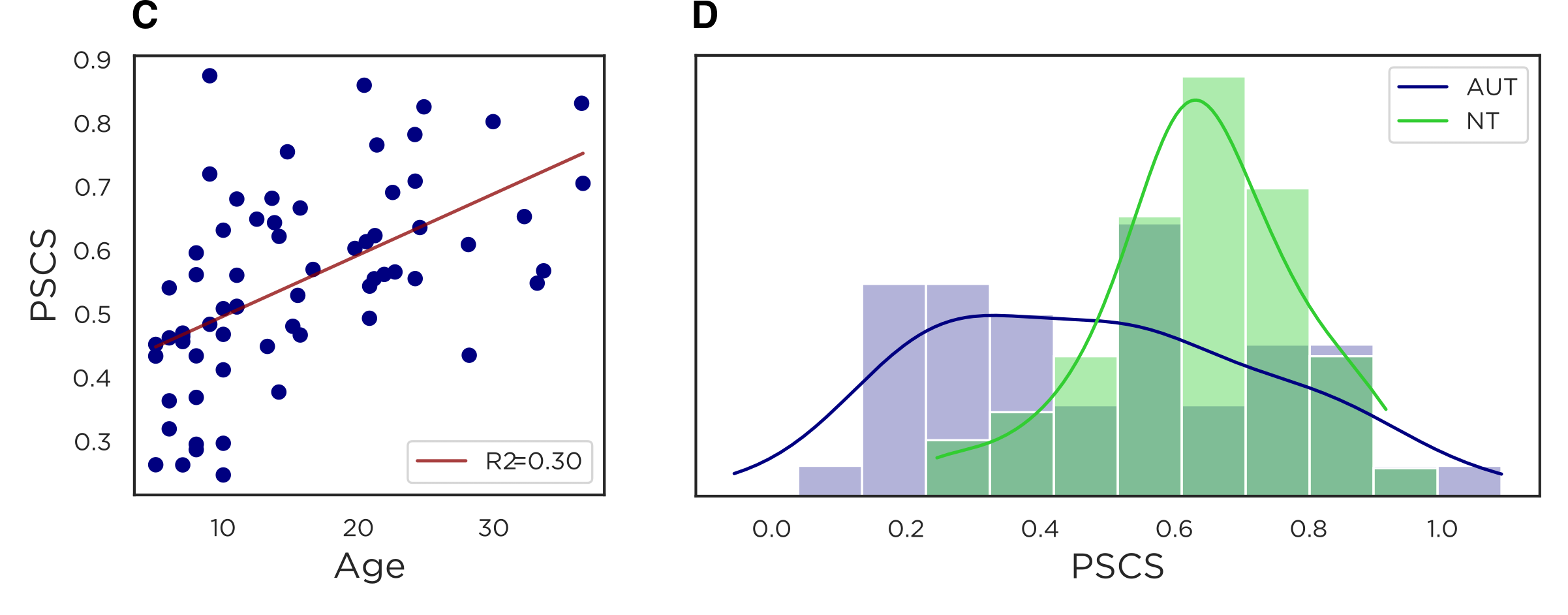} 
\caption{(A) Average per-segment concurrence score (PSCS) per neurotypical (NT) individual versus age. (B) The distributions of PSCS per individual for the autism (AUT) and NT group. }
\label{fig:behavior}
\end{wrapfigure}

\textbf{Behavioral Data}.
Finally, we apply  concurrence to the analysis of facial behavior occurring in a dyadic conversation task. The  behaviors of two conversation partners are expected to be dependent, due to well-established phenomena like nonconscious mimicry~\cite{lakin03} or (nonverbal) backchanneling~\cite{shelley13}. However, quantifying such dependencies has proven challenging, as behavior is captured with multi-dimensional signals (Fig.~\ref{fig:behavior}A,B), and any subset of signals from one conversation partner may depend on the signals of the other partner through an unknown relationship.

We conduct experiments on a dataset of 199 participants (aged 5 to 40 years) engaged in a 3-5-minute semi-structured face-to-face conversation \cite{sariyanidi23}. We quantify social behaviors (i.e., facial expressions and head movements) in each conversation partner with 82-dimensional signals (79 for facial expressions and 3 for head movements) \cite{sariyanidi24}. The concurrence coefficient for $w=4$ seconds is 0.49 (p<0.001), indicating that the behavior signals of the conversation partners are dependent. Moreover, the PSCS allows us to investigate differences within different subsamples. For example, Fig.~\ref{fig:behavior}C shows that the PSCS increases with age (Spearman’s r = 0.61, p < 0.001), indicating that younger school-age children tend to have less behavioral coordination than older children. Additional analyses on a subsample of 12-18 year-olds (N=42) with and without an autism diagnosis (matched on age and sex) indicate that autistic adolescents have reduced coordination with conversation partners relative to neurotypical adolescents (Cohen’s D: 0.8; p =0.003; Fig.~\ref{fig:behavior}D). Together, these results demonstrate that the concurrence method exposes clinically relevant differences in spontaneous social behavioral coordination, without any a priori information about the structure of the coordination.

\section{Limitations of Measuring Dependence with Self-supervision}

While the proposed method showed no propensity to discovering spurious dependencies (Fig.~\ref{fig:snr}), it is not uncommon that signals generated for scientific analyses contain responses to common, extraneous events. These common events render the compared signals statistically dependent, yet since they are extraneous, they are often unrelated to the research question that is being investigated. 

This phenomenon can be observed, for example, in a comparison of EEG signals versus facial behavior. While one may be interested in analyzing these two modalities to uncover potential relationships between mental/emotional states and observable facial behavior, the EEG sensors are sensitive to non-neural as well as neural activities. For example, eyeblinks can generate voltage change that is an order of magnitude larger than cortical activity \cite{plochl12}. While pre-processing algorithms may be able to mitigate the effect of blinking, the downside with a potent dependence detection algorithm is that any residue left from such mitigation efforts will also be detected as statistical dependence.

An alternative approach for accounting extraneous factors is using correction procedures after the dependence is estimated. These procedures are typically used with traditional approaches to measuring statistical dependencies, and their success is yet to be demonstrated for the proposed concurrence algorithm, or alternative algorithms based on (deep) learning, which can expose more nuanced dependencies that can remain on the compared variables even after correction.

\section{Conclusion}
This paper introduced a new approach for measuring statistical dependence, namely, concurrence. We showed that this self-supervised approach can become a standard way of quantifying statistical dependence between time series, as it readily detects a wide range of linear or non-linear dependencies with an off-the-shelf implementation, even from modestly sized samples and noisy data, without requiring empirical (hyper)parameter tuning; and showed no propensity to false discoveries (Type I errors). Future research can further enhance this framework by theoretically establishing the link between statistical dependence and concurrence, while integrating advances in machine learning can ensure that its theoretical potential can be fully actualized. Ensuring that the exposed statistical dependencies are truly of scientific interest remains an open problem.  Therefore the users of concurrence or any other method should verify that the signals do not contain common responses to extraneous factors, or perform pre-processing or post-hoc analyses to ensure that the dependencies that are exposed truly pertain to the research question and phenomena under investigation.


\bibliographystyle{unsrtnat} 
\bibliography{arxiv_version}    


\appendix
\counterwithin{table}{section}                 
\counterwithin{figure}{section}                 
\renewcommand{\thetable}{\thesection.\arabic{table}}
\renewcommand{\thefigure}{\thesection.\arabic{figure}}

\section{Network Architecture, Hyperparameters and Implementation}

The transformations $f(\cdot)$ and $g(\cdot)$ are modelled with separate convolutional neural networks (CNNs), but both CNNs use an identical and well-established architecture. Specifically, each CNN is comprised of $B$ identical blocks concatenated back to back. Each block is comprised of four layers:
\begin{itemize}
    \item Batch normalization layer
    \item Convolutional layer
    \item Dropout
    \item ReLU
\end{itemize}

\noindent The convolutional layer has a stride parameter, which effectively downsamples the signals in time when it is greataer than 1. Moreover, following standard practice, the convolutional layer at each block reduces the number of channels (\ie dimension of signals) by half. 

The training is done by using the Adam optimizer for 100 iterations (Table~\ref{tab:params}), although the code has the option to stop early by using a certain percentage (default 20\%) of the training data as a validation set. During training, we extract four randomly selected segment pairs from each signal pair. Each segment pair is picked to be concurrent (i.e., positive sample) with 50\% probability and non-concurrent also with 50\% probability.

We successfully used the network parameters in Table~\ref{tab:params} to expose dependencies on three types of behavioral signals with divergent characteristics \---fMRI, physiological (ECG and respiration rate) and behavioral data\--- as well as 100 synthesized datasets (Data S1) with a variety of dependence patterns (see Supplementary Text). As such, these parameters have proven capability to expose a wide range of non-linear dependencies. If the two signals are dependent but through a large temporal lag, one may need to increase the number of blocks $B$. 

The number of output channels of the (first) convolution layer, $C$, controls the complexity of functions that can be modeled with $f$ or $g$\----the higher the $C$ the more complex functions can be modeled. We set this parameter to $C=512$, which worked successfully across a wide range of dependence patterns\----from simple linear dependence between one-dimensional input segments to complex non-linear dependencies, including with multi-dimensional signals, such as in our experiments with behavioral data (82 dimensions). In other words, we did not observe any harm in setting this parameter to higher values than needed (\eg linear dependence between one-dimensional signals could have been technically be exposed with $C=1$ and $B=1$). In cases where one deals with very high-dimensional input signals (\eg $M_1$ or $M_2$ in the order of hundreds or thousands) or one expects a very complex dependence, one may need to set $C$ to higher values than $512$. Also, in cases where computational efficiency is a priority, one may reduce this value possibly without harm, since $C=512$ is possibly a larger value than needed in many applications.



\begin{table}
\centering
\begin{tabular}{|l|c|} \hline
Kernel size of first conv. layer & $5$ \\
Kernel size of other conv. layers & $3$ \\
Step size (stride) at conv. first layer & $3^*$ \\
Step size (stride) at other conv. layers & $2^*$ \\
Number of blocks ($B$) & $3$ \\ 
Number of output channels at 1st conv. layer  ($C$) & $512$\\
Number of output channels at $b$th conv. layer  & $512/2^{(b-1)}$\\
Dropout rate & $0.25$ \\ 
Optimizer & Adam${}^{\text{51}}$ \\ 
Number of iterations &  $100$ \\ 
Learning rate &  $10^{-4}$ \\ \hline
\end{tabular}
\caption{Parameters of the CNNs that we use. ${}^{*}$The step sizes larger than $1$ effectively downsample the input, and if the segment size $w$ is too small, the output of a convolutional kernel may be empty. To avoid the latter, one may need to reduce the stride sizes accordingly.}
\label{tab:params}
\end{table}


\newpage

\begin{figure}
\centering
\includegraphics[scale=0.39]{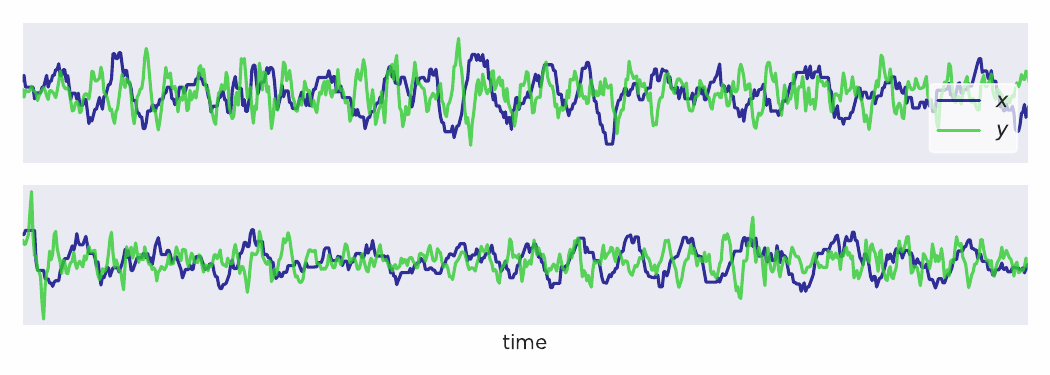} 
\includegraphics[scale=0.39]{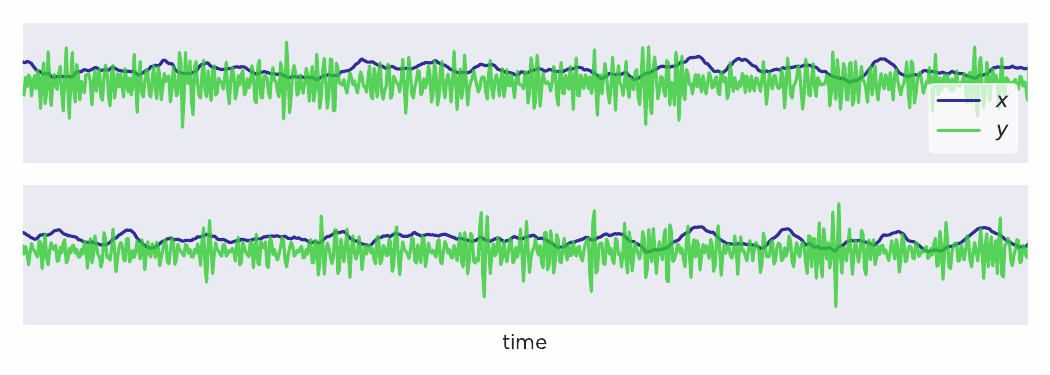} \\
\includegraphics[scale=0.39]{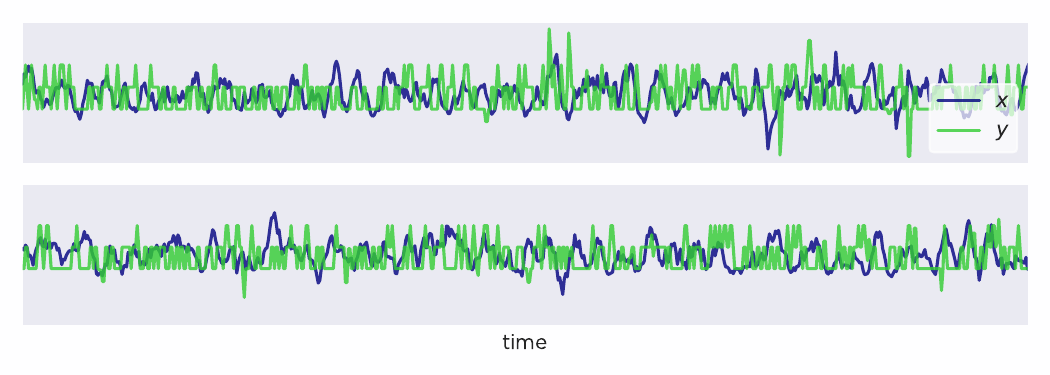} 
\includegraphics[scale=0.39]{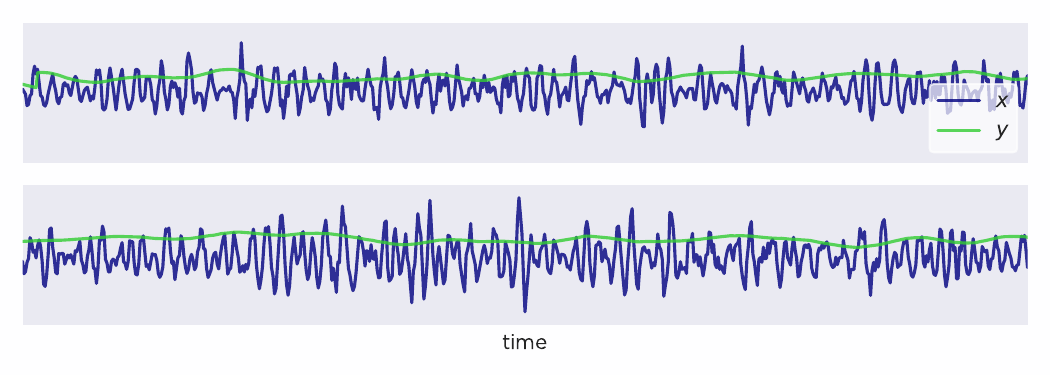} \\
\includegraphics[scale=0.39]{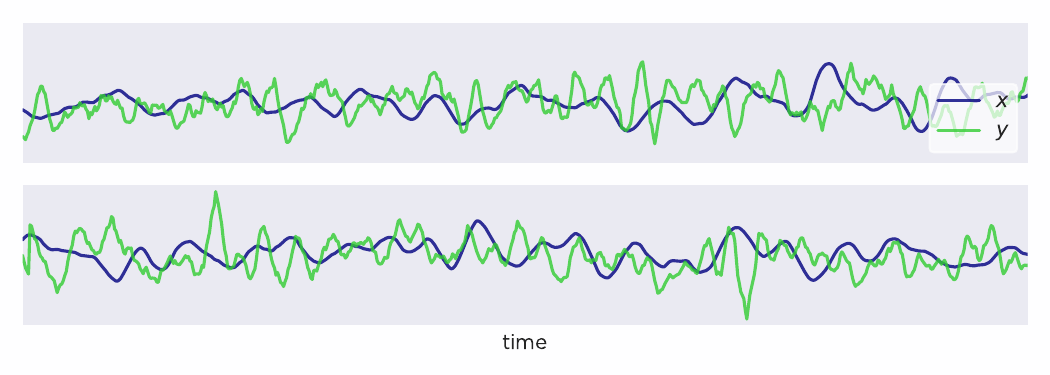} 
\includegraphics[scale=0.39]{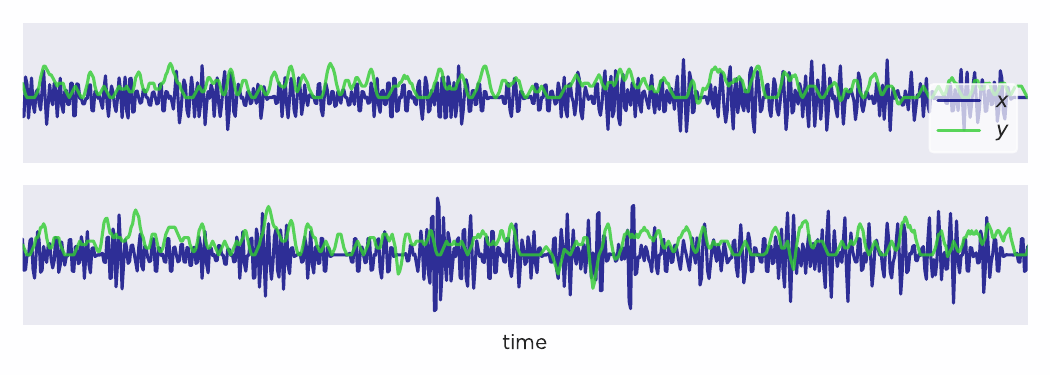} \\
\includegraphics[scale=0.39]{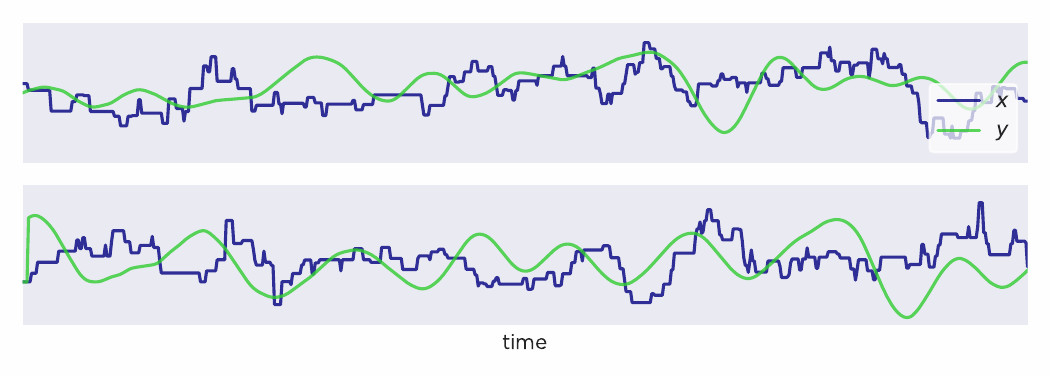} 
\includegraphics[scale=0.39]{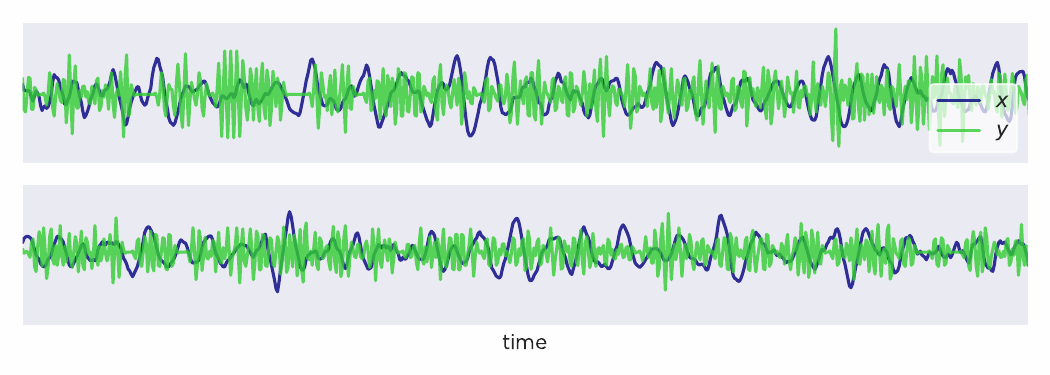} \\
\includegraphics[scale=0.39]{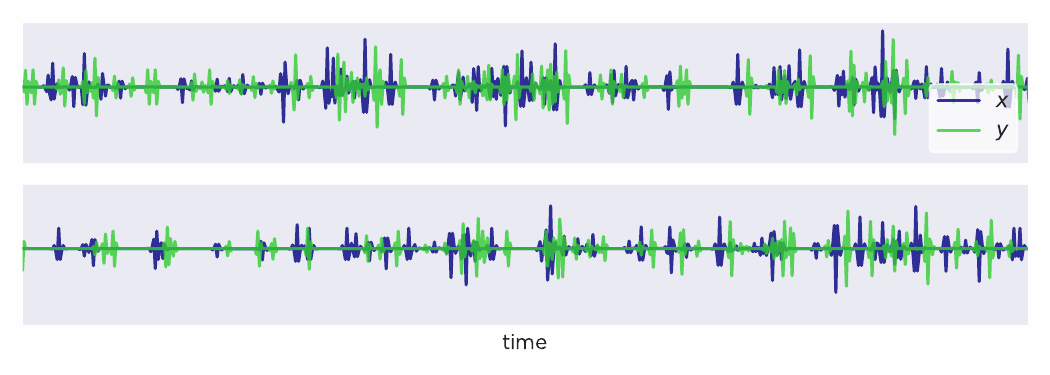} 
\includegraphics[scale=0.39]{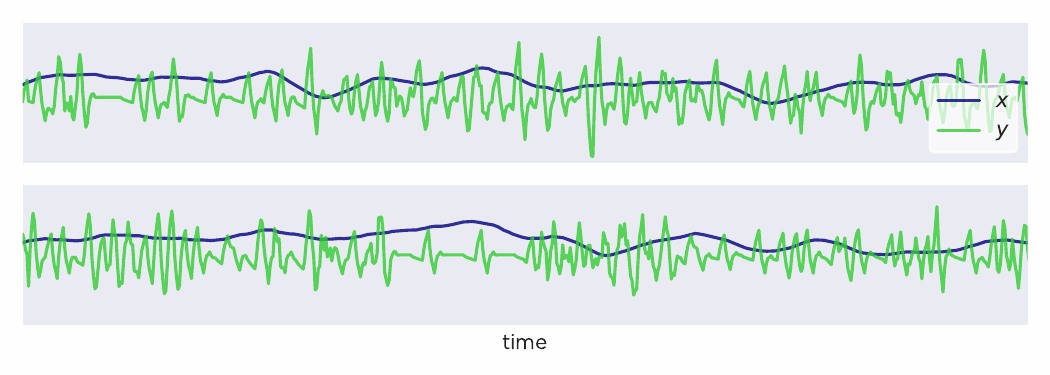} \\
\includegraphics[scale=0.39]{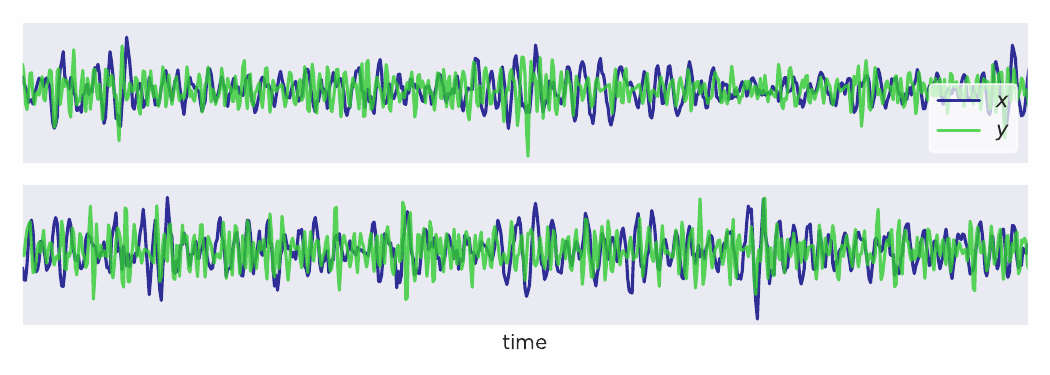} 
\includegraphics[scale=0.39]{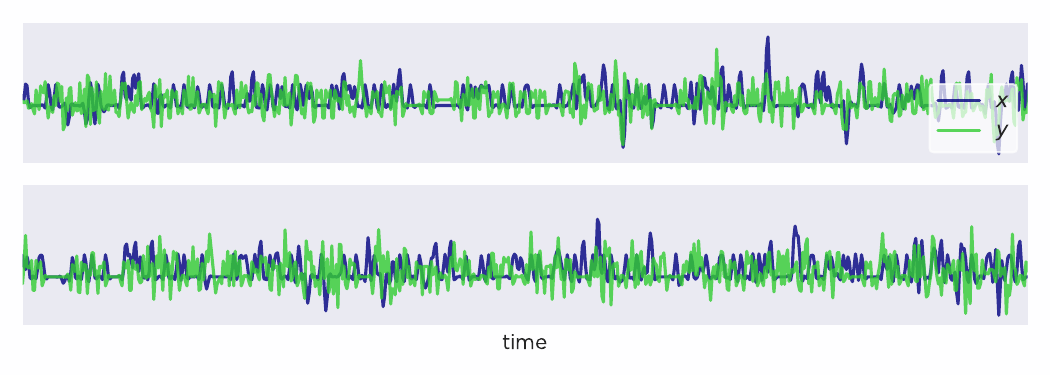}
\caption{\textbf{Pairs of signals representative from 12 out of the 100 synthesized datasets.} All illustrated signal pairs are dependent, and they are generated as described in Supplementary Text. The raw data for all 100 synthesized datasets is provided in Data S1.}
\label{fig:signals}
\end{figure}

\newpage

\begin{figure}
\centering
\includegraphics[scale=0.31]{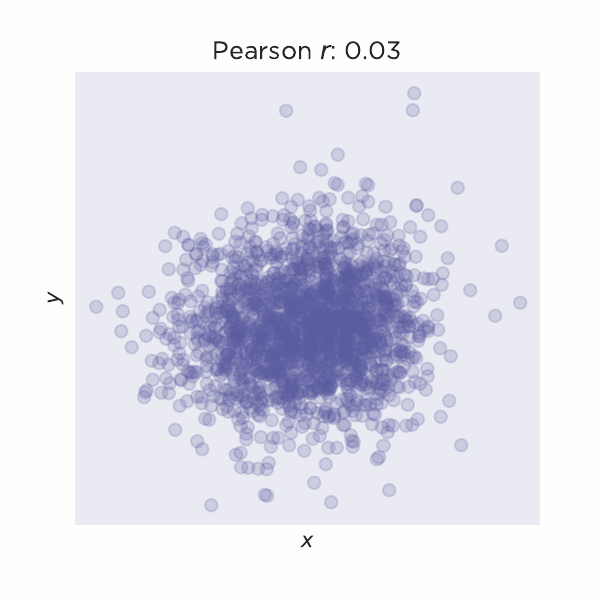} 
\includegraphics[scale=0.31]{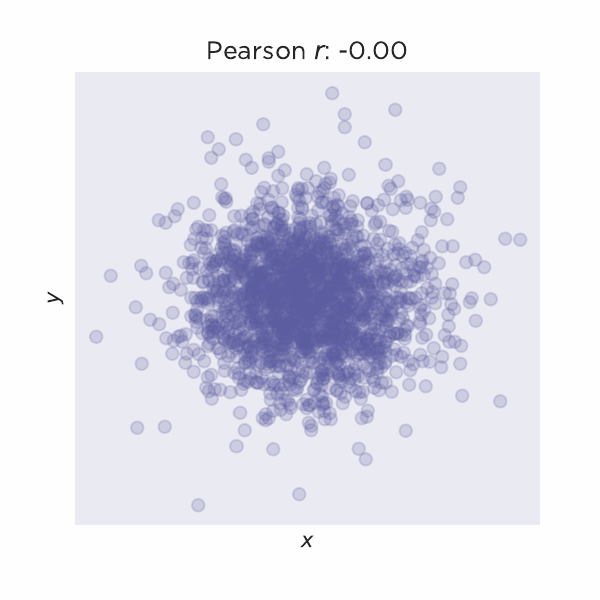} 
\includegraphics[scale=0.31]{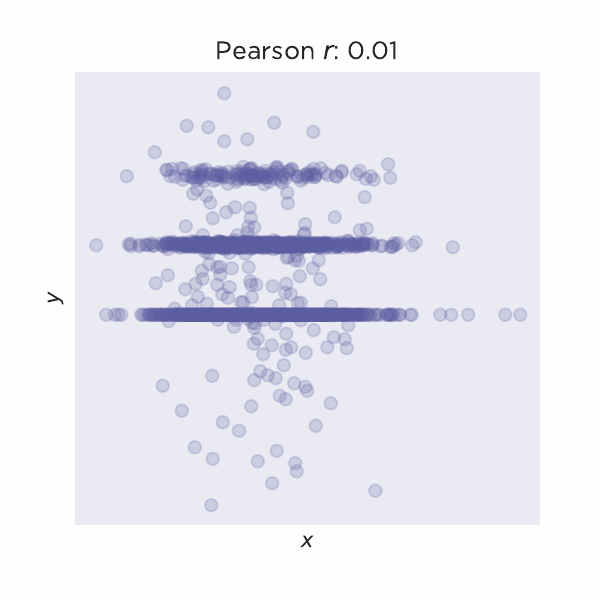} 
\includegraphics[scale=0.31]{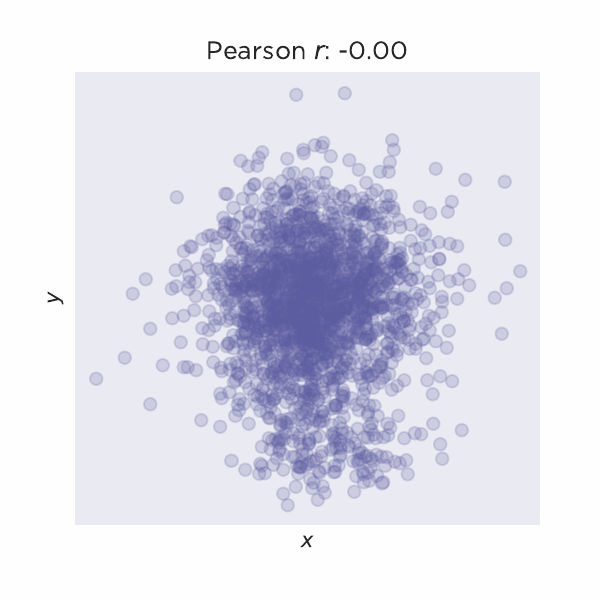} \\
\includegraphics[scale=0.31]{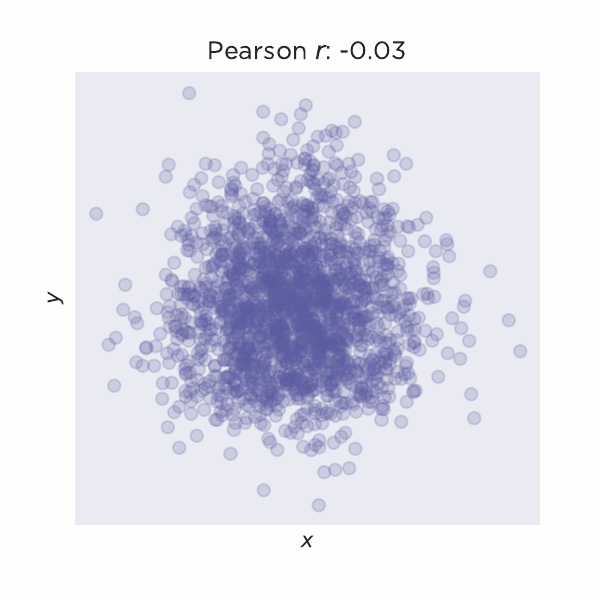} 
\includegraphics[scale=0.31]{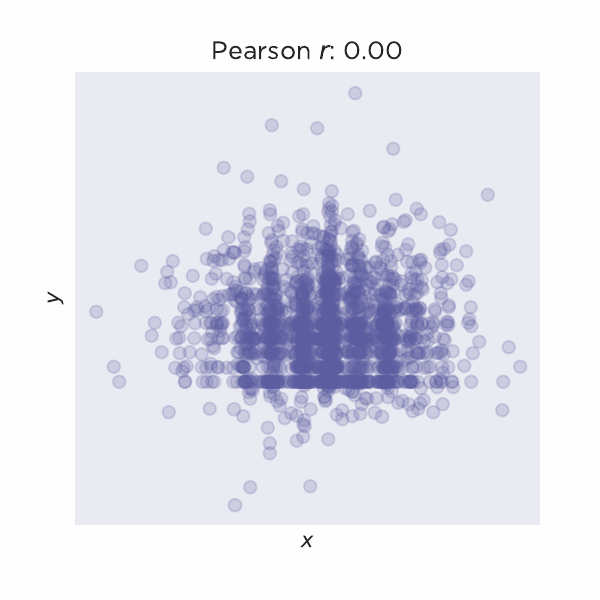} 
\includegraphics[scale=0.31]{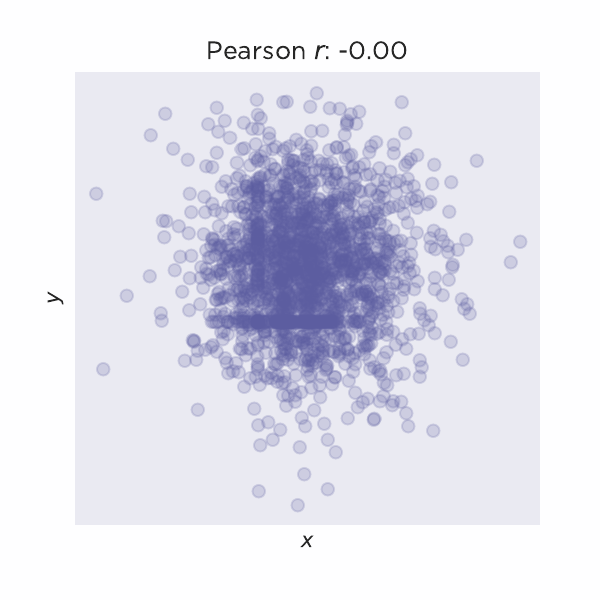} 
\includegraphics[scale=0.31]{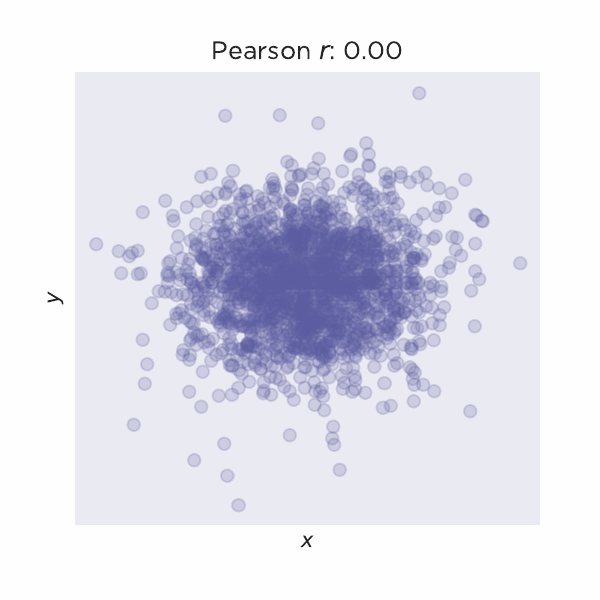} \\
\includegraphics[scale=0.31]{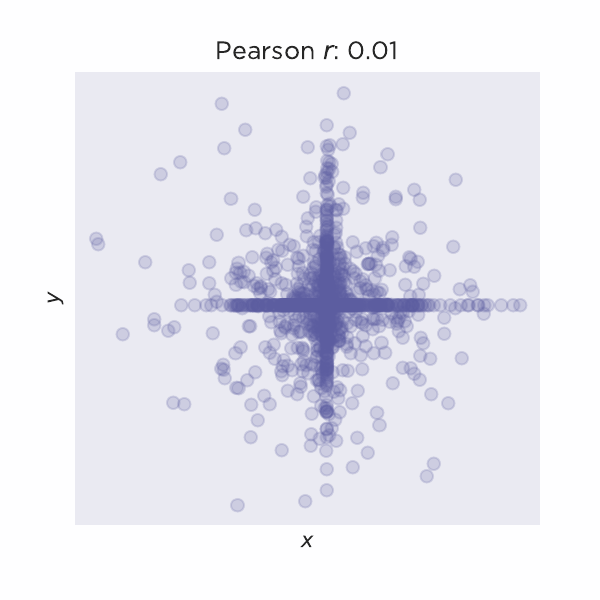} 
\includegraphics[scale=0.31]{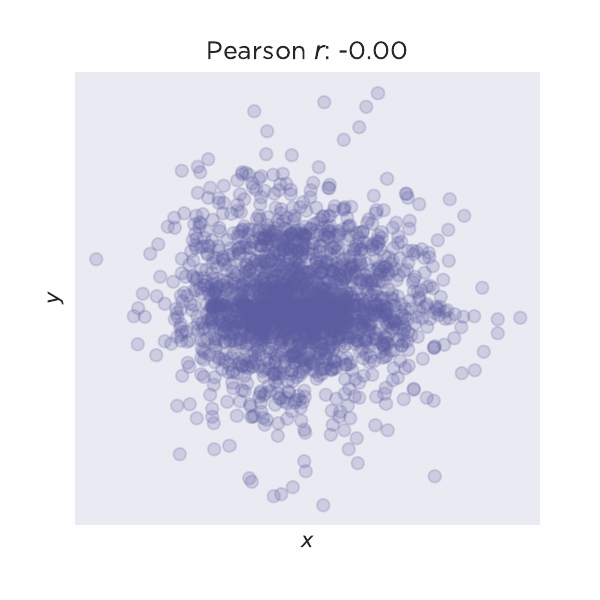} 
\includegraphics[scale=0.31]{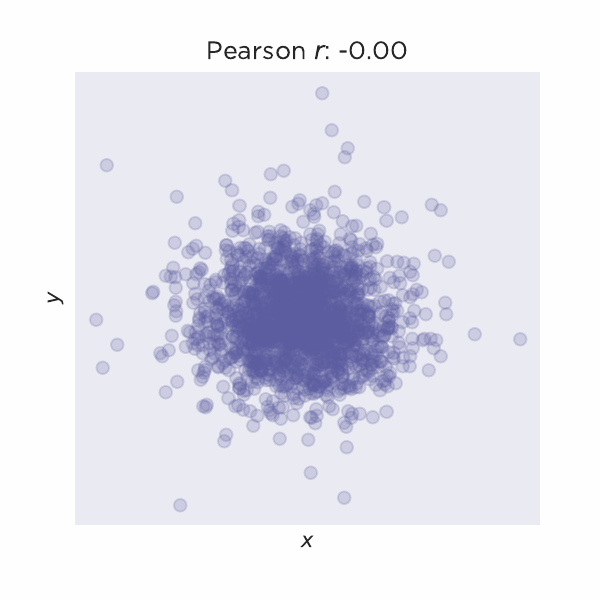} 
\includegraphics[scale=0.31]{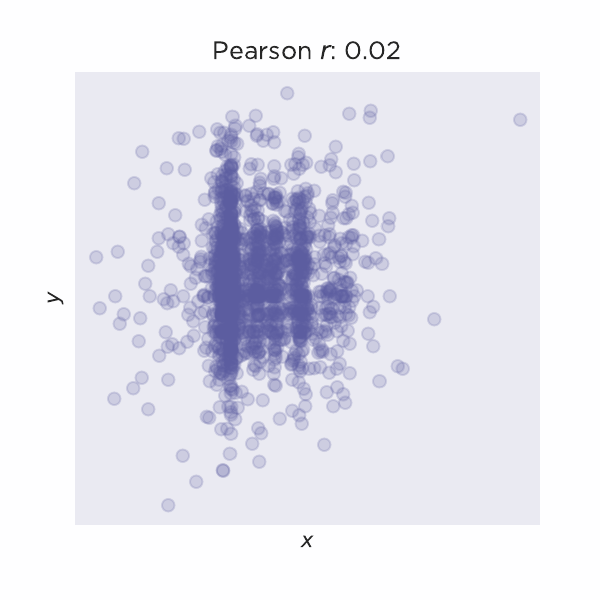}
\caption{\textbf{The scatter plot and correlation between the signal pairs illustrated in Fig.~\ref{fig:signals}).} The correlation (Pearson's $r$) between is almost always approx. 0, indicating non-linear dependence.}
\label{fig:corrs}
\end{figure}

\end{document}